\documentclass[graybox,pdftex]{svmult}

\usepackage{mathptmx}       
\usepackage{helvet}         
\usepackage{courier}        
\usepackage{type1cm}        
\usepackage{makeidx}         
\usepackage{graphicx}        
\usepackage{multicol}        
\usepackage[bottom]{footmisc}

\usepackage{fancyhdr}

\makeindex             

\begin{document}

\title*{Computational Steering of Complex Flow Simulations}
\author{Atanas Atanasov\inst{1}, Hans-Joachim Bungartz\inst{1} \and
	J\'{e}r\^{o}me Frisch\inst{2} \and Miriam Mehl\inst{1} \and Ralf-Peter
	Mundani\inst{2} \and Ernst Rank\inst{2} \and Christoph van
	Treeck\inst{3}}

\authorrunning{A.~Atanasov, H.-J.~Bungartz, J.~Frisch, M.~Mehl, R.-P.~Mundani, E.~Rank, C.~van Treeck}

\institute{Department of Informatics, Technische Universit\"at M\"unchen \email{\{atanasoa,bungartz,mehl\}@in.tum.de}
	\and Chair for Computation in Engineering, Technische Universit\"at M\"unchen \email{\{frisch,mundani,rank\}@bv.tum.de}
	\and Fraunhofer-Institut f\"ur Bauphysik, Holzkirchen \email{treeck@ibp.fraunhofer.de}}

%
%
\maketitle

\abstract{Computational Steering, the combination of a simulation back-end with
a visualisation front-end, offers great possibilities to exploit and optimise
scenarios in engineering applications. Due to its interactivity, it requires
fast grid generation, simulation, and visualisation and, therefore, mostly has
to rely on coarse and inaccurate simulations typically performed on rather
small interactive computing facilities and not on much more powerful
high-performance computing architectures operated in batch-mode. This paper
presents a steering environment that intends to bring these two worlds -- the
interactive and the classical HPC world -- together in an integrated way. The
environment consists of efficient fluid dynamics simulation codes and a
steering and visualisation framework providing a user interface, communication
methods for distributed steering, and parallel visualisation tools. The gap
between steering and HPC is bridged by a hierarchical approach that performs
fast interactive simulations for many scenario variants increasing the accuracy
via hierarchical refinements in dependence of the time the user wants to wait.
Finally, the user can trigger large simulations for selected setups on an HPC
architecture exploiting the pre-computations already done on the interactive
system.}

\pagestyle{empty}
\thispagestyle{fancy}
\lhead{}
\chead{}
\rhead{}
\lfoot{\scriptsize This is a pre-print of an article published in Wagner~S., Steinmetz~M., Bode~A., M\"uller~M.\ (eds) High Performance Computing in Science and Engineering, Garching/Munich 2009. The final authenticated version is available online at: https://doi.org/10.1007/978-3-642-13872-0\_6}
\cfoot{}
\rfoot{}

\flushbottom

%
%
\section{Introduction}
Computational Steering is the coupling of a simulation back-end with a
visualisation front-end in order to interactively exploit design alternatives
and/or optimise (material) parameters and shape. Therefore, different aspects
such as grid generation, efficient algorithms and data structures, code
optimisation, and parallel computing play a dominant role to provide quick
results (i.\,e.\ several simulation and visualisation updates per second in
case of modifications of the underlying data) to keep up the principle of cause
and effect, which is necessary to gain better insight and a deeper
understanding of problems from the field of engineering applications.
Nevertheless, even nowadays interactivity and high-performance computing (HPC)
are still a contradiction, as most HPC systems do not provide interactive
access to the hardware.

As a remedy for the latter one, a two-stage approach (i.\,e.\ interactive
pre-processing of ``low level'' problems and parallel processing of ``high
level'' problems) helps to bridge the gap between small -- and typically
interactive -- systems for a quick quantitative analysis and large -- and
typically batch -- HPC systems for a complex qualitative analysis. Such an
approach also provides the advantage of reducing the amount of long and, thus,
expensive simulation runs to those necessary only without waisting additional
computing time for redundant computations. To ensure a \emph{seamless}
transition from ``low level'' problems on coarse grids with few thousands of
unknowns to ``high level'' problems on fine grids with many millions of
unknowns, hierarchical approaches are indispensable.

This also has a significant relevance for the practical usage of computational
steering and HPC in industrial applications, as most approaches there suffer
from a insufficient integration of HPC into the workflow of industrial processes.
Hence, from the very beginning one of our main objectives was to provide a
framework for engineering applications that not only addresses challenging
mathematical and computer science related questions, but also combines and
consolidates the two conflicting aspects of interactivity and high-performance
computing. Therefore, we will show the benefits of our framework for the
interactive control of different engineering applications running on parallel
architectures.

The remainder of this paper is as follows. Section~\ref{sec:steering} presents
the ingredients of the steering environment. As this environment does not cover
the whole range of applicability of the underlying approaches,
Sect.~\ref{sec:others} describes two further applications that have been or
will be coupled to the steering and visualisation framework. Finally, we draw
a short conclusion and give an outlook on the future work in
Sect.~\ref{sec:summary}.

%
%
\section{Computational Steering Environment}
\label{sec:steering}
In order to increase the performance, i.\,e.\ decrease simulation and
visualisation response time of our steering environment as well as to prepare
it for a later HPC usage, several measures have been taken. This was done with
a straight focus on the two-way approach as described above, where small
systems are used for an interactive data exploration before a 
high-quality analysis (based on the parameters explored) is launched as
(massively) parallel job on large HPC systems.

\subsection{Hierarchical Approach} 
The main idea in joining the interactively computed small systems with the
large parallel systems computed on HPC architectures is to exploit hierarchies
of grid levels or discretisation orders. As a response on each user input, a
simulation on a very coarse grid or with lowest discretisation order is
triggered such that first visualised results are available very fast. Depending
on the time given -- that is the time the user wants to wait for more accurate
results -- the simulation is refined in a recursive manner. Each of these
refinement steps adds a new layer of grid points to decrease the mesh width or
additional degrees of freedom at existing grid points to enhance the
approximation order. This allows to quickly check results for numerous input
configurations, to examine those that seem to be relevant more accurately and,
finally, to start large HPC simulations only for a few scenarios of particular
interest. Hereby the refined simulations already profit from the coarser ones
in a full multigrid manner. Codes such as \textsf{iFluids}, \textsf{Peano}, and
the \emph{p}-FEM structural mechanics codes described below naturally fit this
approach as they inherently already provide the required hierarchy.  

\subsection{iFluids}
The kernel of our steering framework is a Lattice-Boltzmann fluid solver which
has been developed by our group and ported to the former HPC system -- the
pseudo-vector computer Hitachi SR8000-F1 -- installed at Leibniz-Rechenzentrum
(LRZ). This fluid solver -- called \textsf{iFluids} \cite{Treeck:Bph:07} -- was
running interactively on the Hitachi while coupled with the interactive
visualisation nodes also available at LRZ for computational steering
applications. Due to the replacement of the old HPC system with the SGI Altix
4700 severe changes of \textsf{iFluids} became necessary in order to run it
successfully on the new system.  These changes comprise to switch from a pure
MPI-based implementation to a cache-efficient hybrid approach (MPI/OpenMP) to
benefit also from the Itanium CPUs' local shared memory as well as to modifiy
the communication and data distribution pattern, such that it optimally suits
the underlying network topology (2D tori connected via a fat tree) in order to
minimise latency.

\begin{figure}[h]
\begin{minipage}{0.49\textwidth}
\begin{center}
\includegraphics[width=0.7\textwidth, angle=-90]{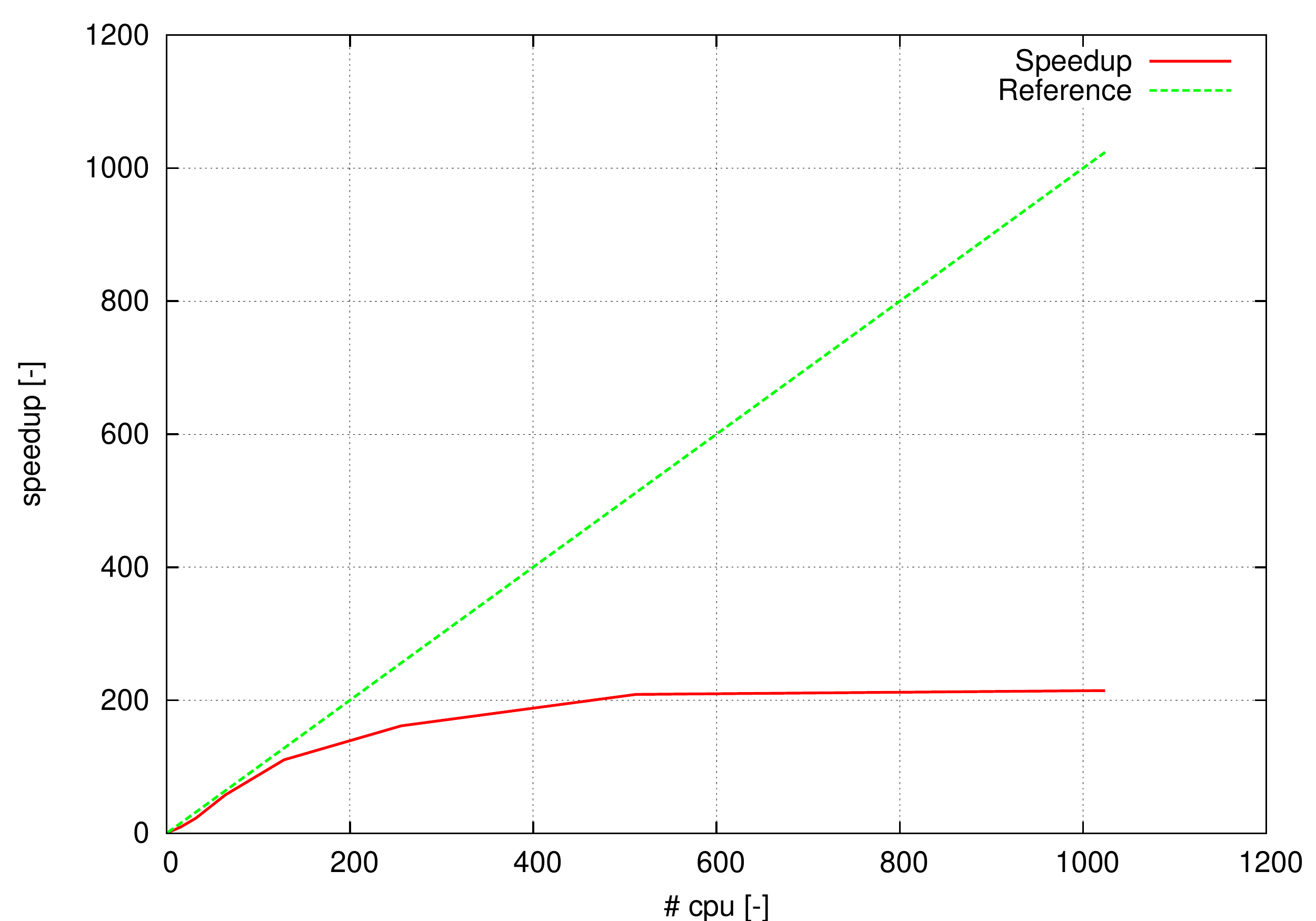}
\end{center}
\end{minipage}
\begin{minipage}{0.49\textwidth}
\begin{center}
\includegraphics[width=0.9\textwidth, angle=0]{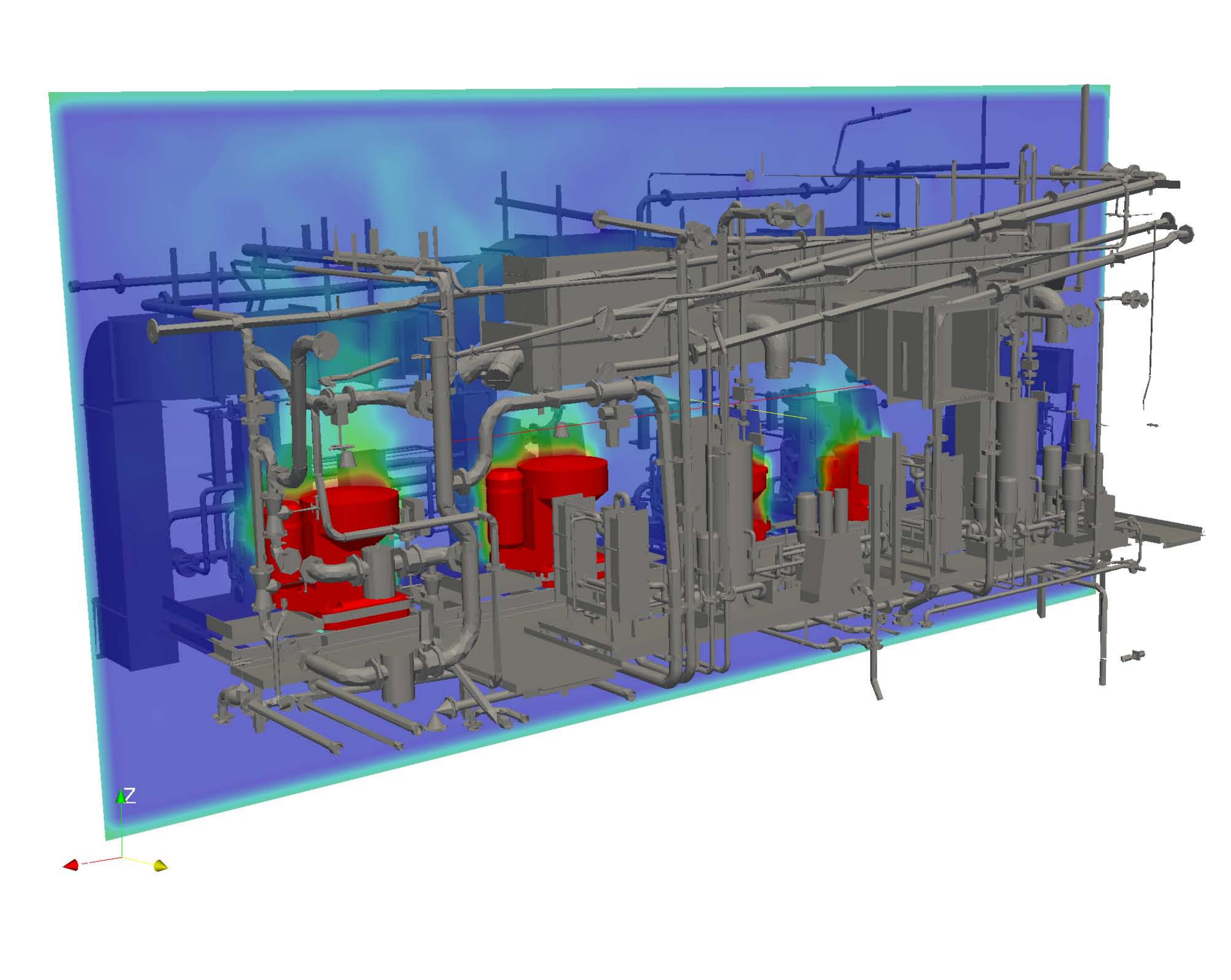}
\end{center}
\end{minipage}
\caption{Performance results running \textsf{iFluids} on the SGI Altix 4700 --
	 left-hand side shows strong speedup values, right-hand side shows
	 simulation results for a complex geometry. }
\label{speedup}
\end{figure}

As the porting of \textsf{iFluids} is still work in progress, current
performance measurements (up to 1024 processes) on the Altix do not yet reveal
the full potential of the parallel code, nevertheless already sound very
promising. For a problem size with 7.5 million degrees of freedom a nearly
linear speedup (strong scaling) up to $p = 64$ processes could be observed
which strongly drops for growing numbers of \emph{p} (see Fig.~\ref{speedup}).

Further investigations on this behaviour showed that the major drawback of the
current parallelisation is the regular block decomposition of the domain that
leads to partitions consisting of mostly or entirely obstacle cells only, for
which no computation have to be performed. This leads to an unbalanced load
situation. Therefore and due to the frequent geometry and refinement depth 
changes in a steering environment, a more enhanced adaptive and dynamical
load balancing strategy is inevitable. A modified master-slave concept which 
has been developed by our group (see next section) is being incorporated into
\textsf{iFluids} at the moment.

\subsection{Adaptive Load Balancing}
Within a related project for structural analysis using the \emph{p}-version
finite element method (\emph{p}-FEM, \cite{Duester:pFEM}) -- i.\,e.\ increasing
the polynomial degree \emph{p} of the shape functions for better accuracy
without changing the discretisation -- a similar behaviour regarding unbalanced
load situations has been observed when using a hierarchical approach (octrees)
for domain decomposition \cite{Mundani:mpi}. Therefore, we have implemented an
adaptive load balancing strategy based on the idea of task stealing---a
modified master-slave concept, that takes into account varying workload on the
grid nodes.

Here, a master process first analyses the tree and estimates the total amount
of work (measured in floating-point operations) per node.  In the next step,
those nodes are assigned to processes called \emph{traders} -- an intermediate
layer between master and slaves -- to prevent communication bottlenecks in the
master and, thus, making this approach also scaleable for large amounts of
processes. The traders define tasks (i.\,e.\ systems of linear equations for
domain partitions), ``advertise'' them via the master to the slaves, and take
care about the corresponding data transfer. They also keep track about
dependencies between the tasks and update those dependencies with each result
sent back from a slave. Benchmark computations with different ratios of traders
and slaves have shown good results with respect to the average percentage a
single slave is busy during the entire runtime. This is important to obtain
high update rates in case of frequent re-computations which are necessary for
interactive computational steering applications.

Hence, \textsf{iFluids} can also benefit from this approach. By applying a
hierarchical organisation of the computational domain, the master process could
easily identify regions mostly consisting of obstacle cells when doing its work
load estimation. Such a region could then be combined with neighbouring regions
to a larger task which is processed by a single slave to achieve a better
computation-communication-ratio. As this is still work in progress, there are
no current results so far.

\subsection{Remote Visualisation and Steering Framework}
\begin{figure}[h]
\begin{center}
\includegraphics[width=0.8\textwidth]{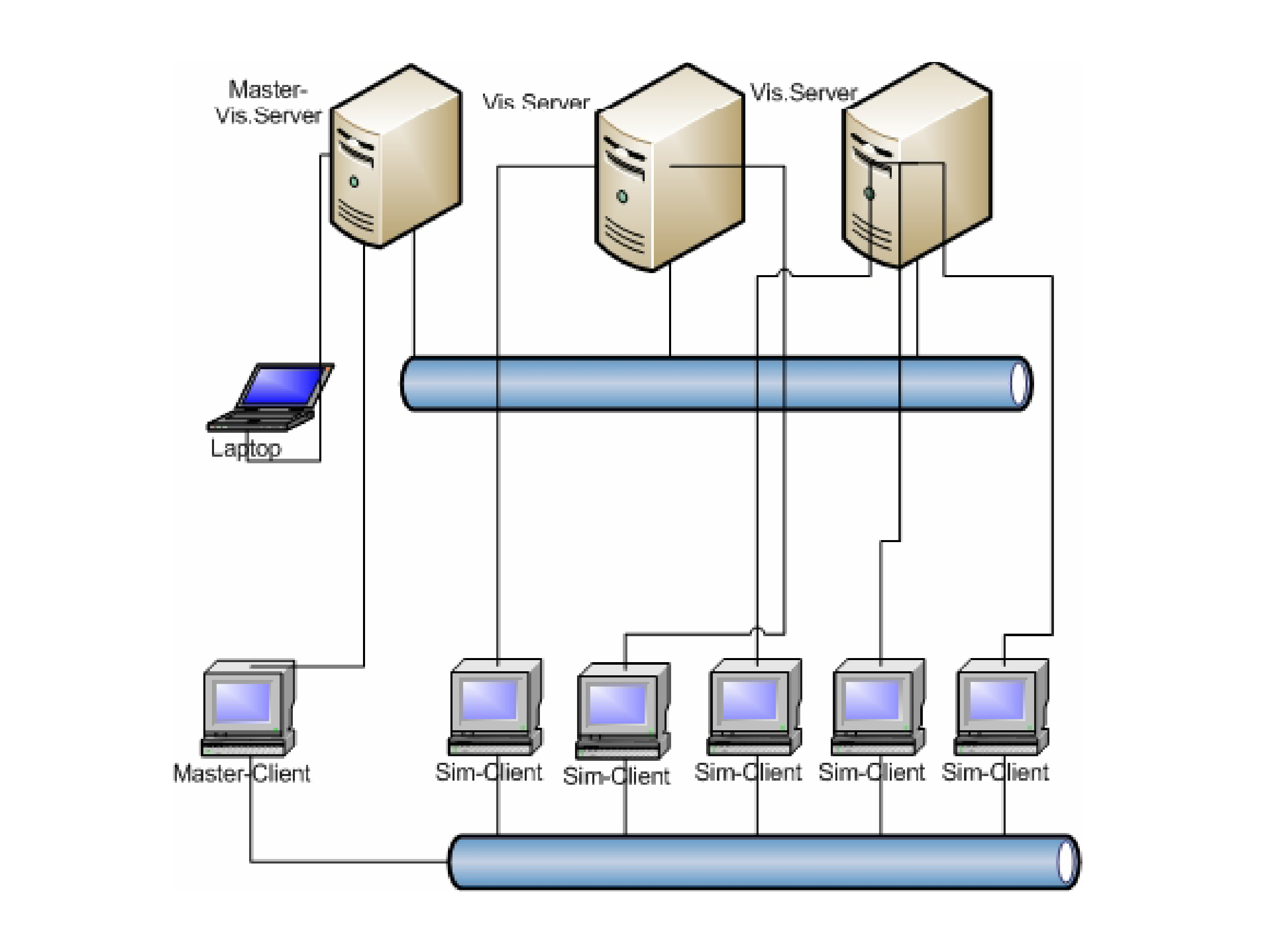}
\end{center}
\caption{\label{fig:atanas_01} Steering environment with parallel remote
	 visualisation and parallel simulation (taken from
	 \cite{Atanasov:09:DA}).}
\end{figure}

For fast visualisation and user interaction, a remote and parallel
visualisation and steering framework has been developed in
\cite{Atanasov:09:DA}. It is based on the idea of a distributed application.
That is, the steering and visualisation application, the underlying simulation,
and the user interface run on separate computing facilties. The interaction
between these components is realised via remote procedure calls (RCP) and TCP
sockets.  As our task is to bring together interactive simulations and
visualisations with HPC applications, i.\,e.\ large systems of equations to be
solved and large data sets to be visualised, the visualisation and simulation
are parallel processes themselves as displayed in Fig.~\ref{fig:atanas_01}. 
 
The visualisation is based on the Visualization Toolkit (VTK,
\cite{VTK:3rdEdition,VTK:www}). For scalar data sets, it provides a colour
mapping as well as iso-lines or iso-surfaces enhanced by cutting planes that
can be displaced and rotated interactively. Vector data such as flow velocities
are visualised using streamlines, dashed streamlines with glyphs, or
streambands. Geometries are represented by surface triangulations and a
bounding box widget that allows to scale, displace, or rotate the geometry. 

\begin{figure}[h]
\begin{center}
\includegraphics[width=0.95\textwidth]{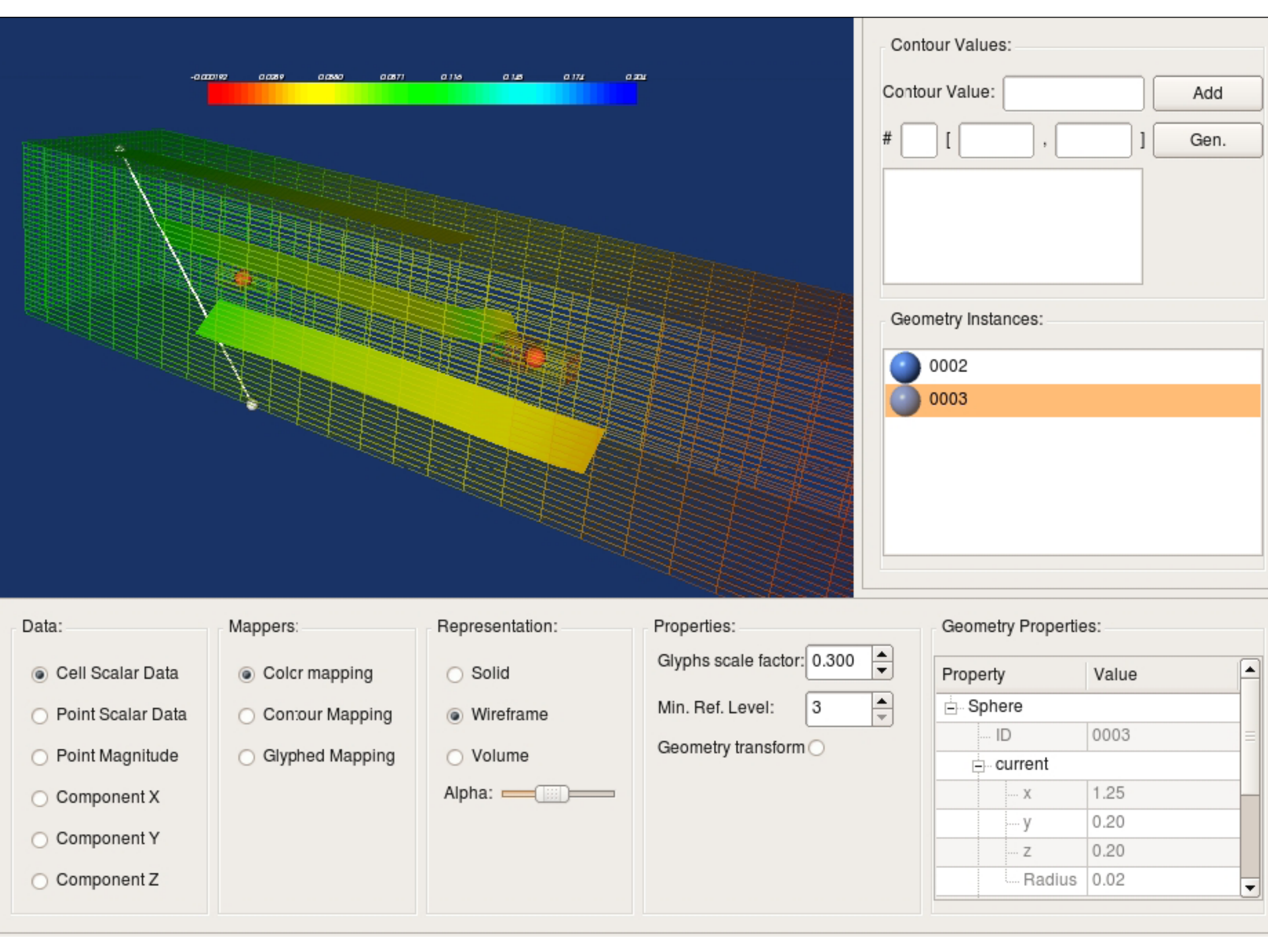}
\end{center}
\caption{\label{fig:atanas_02} Steering User Interface with a streamband
	 visualisation for a flow computed with the \textsf{Peano} CFD solver.
	 The scenario is a channel flow with a spherical obstacle. The second
	 sphere has been added at runtime. At the right bottom, properties of
	 the geometry are displayed (taken from \cite{Atanasov:09:DA}).} 
\end{figure}

The user interface consists of a 3D-viewer, a geometry catalogue, a geometry
browser, and a control panel. It allows the user to change geometries (add,
delete, move, or scale geometrical objects), choose data to be visualised
(velocities or pressure, e.\,g.), select visualisation techniques (streamlines
or streambands, e.\,g.), and to examine simulation results from different views
and with different techniques. Figure~\ref{fig:atanas_02} shows a screenshot of
the user interface with a visualisation of a fluid dynamics scenario.
 
The visualisation is parallelised following a data parallel approach.
Visualisations are performed in parallel for subdomains of the entire scenario.
The bottleneck of this approach is the composition of all subdomain pictures to
a picture of the entire scenario at the end of the visualisation process. A
binary space partition (BSP) tree approach avoids the accumulation of the whole
composition work in one master process.  It recursively joins pictures
associated to the same father in a bottom-up traversal of the BSP tree.

\begin{figure}[h]
\begin{minipage}{0.49\textwidth}
\begin{center}
\includegraphics[width=0.9\textwidth]{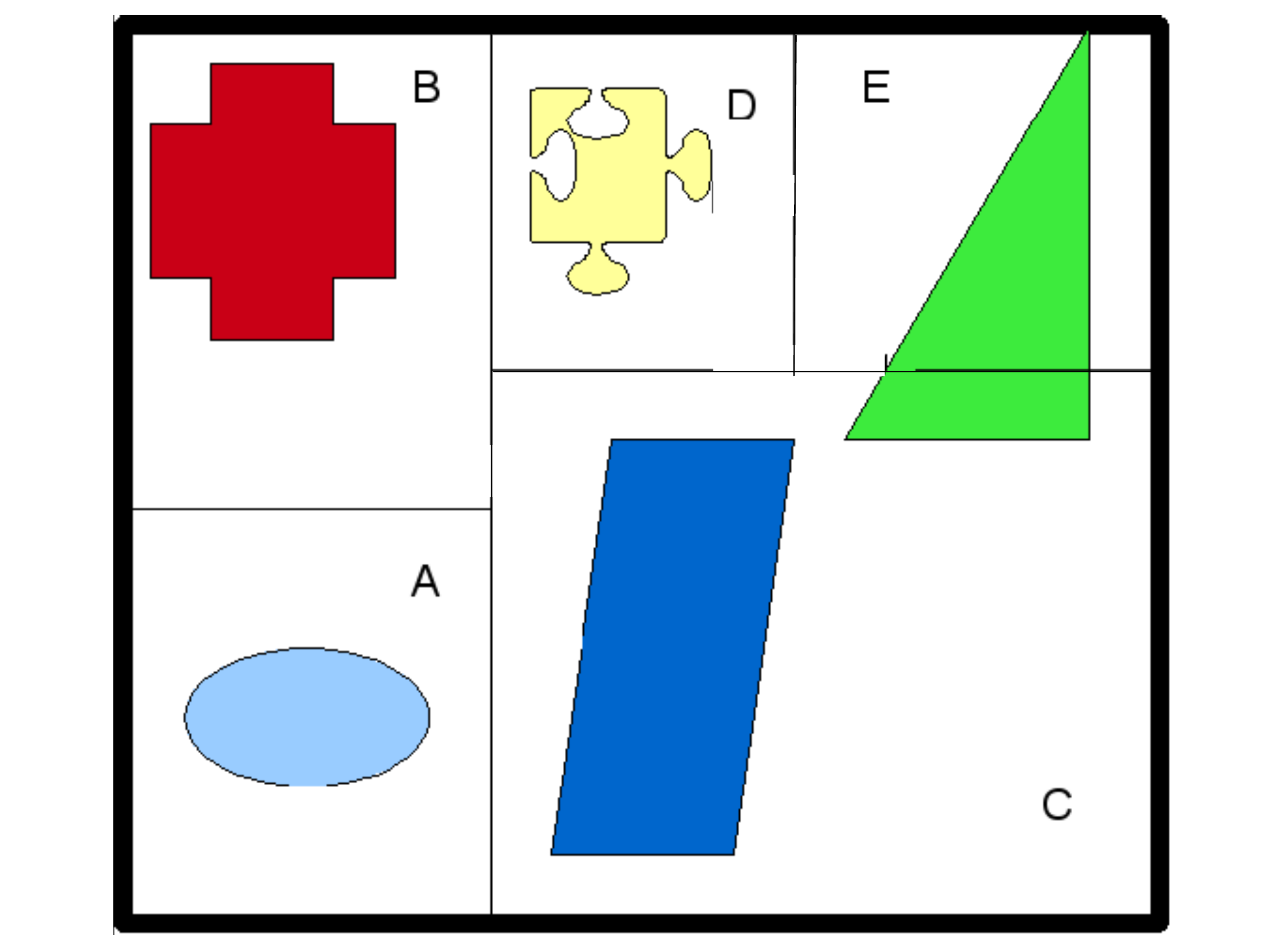}

{\bf (a)}
\end{center}
\end{minipage}
\begin{minipage}{0.49\textwidth}
\begin{center}
\includegraphics[width=0.9\textwidth]{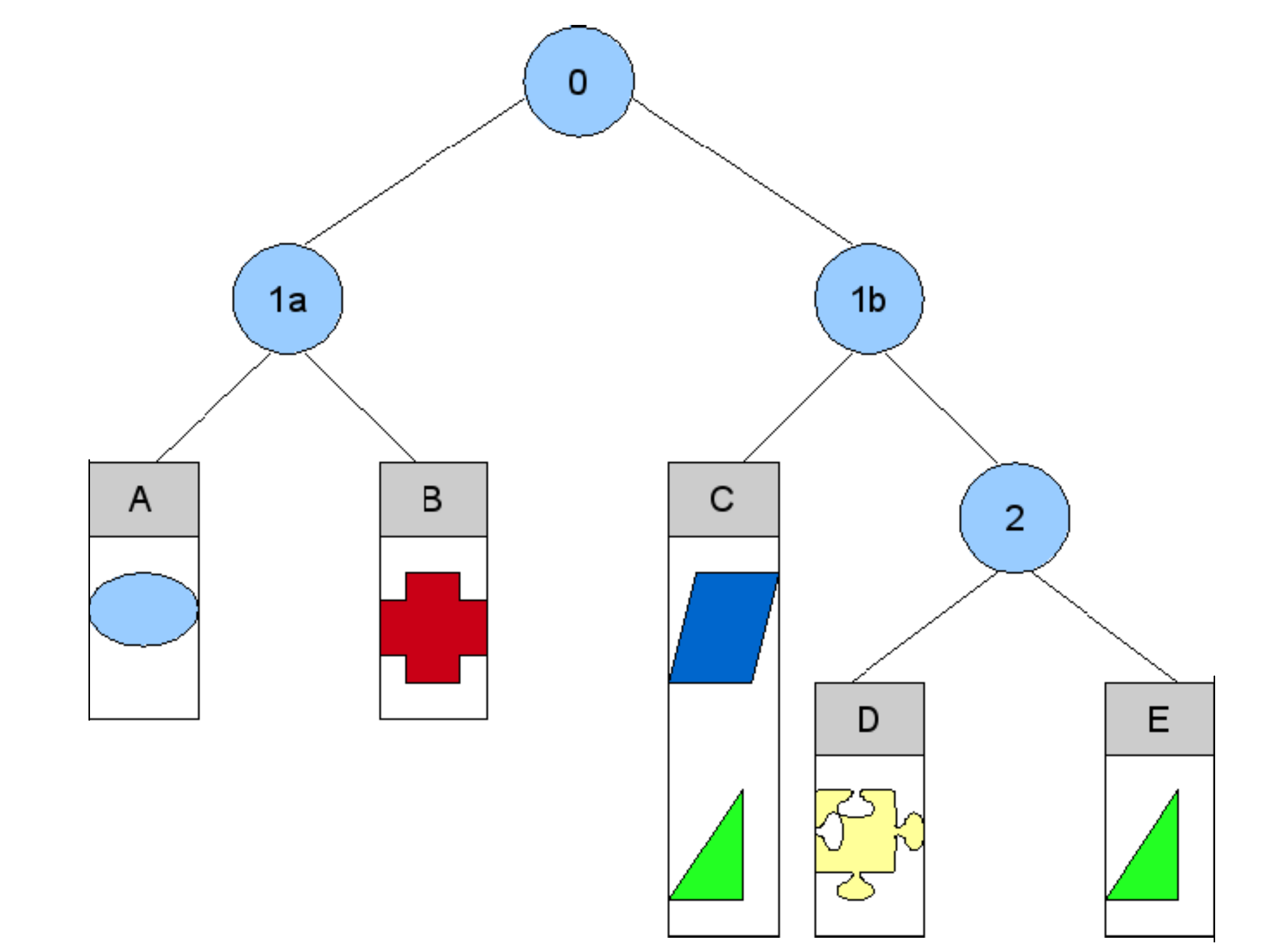}

{\bf (b)}
\end{center}
\end{minipage}
\caption{\label{fig:atanas_04} Example for a BSP-tree domain partitioning for
	 parallel visualisation (taken from \cite{Atanasov:09:DA}). (a) Spatial
	 decomposition according to the BSP-tree; (b) BSP-tree and data
	 structure for this example.} 
\end{figure}

Figure~\ref{fig:atanas_04} shows an example of a domain splitting using a BSP
tree. In this example, the subdomains $D$ and $E$ would be joined first to a
larger domain $DE$. In a second step, $A$ and $B$ would be joined to $AB$. In
parallel, $DE$ would be joined with $C$ to $CDE$, and, finally, $AB$ and $CDE$
would be joined to the entire scenario. In our applications such as Peano, we
use a particular form of BSP trees -- octree-like space-partitioning trees.

\begin{figure}[h]
\begin{minipage}{0.49\textwidth}
\begin{center}
\includegraphics[width=\textwidth]{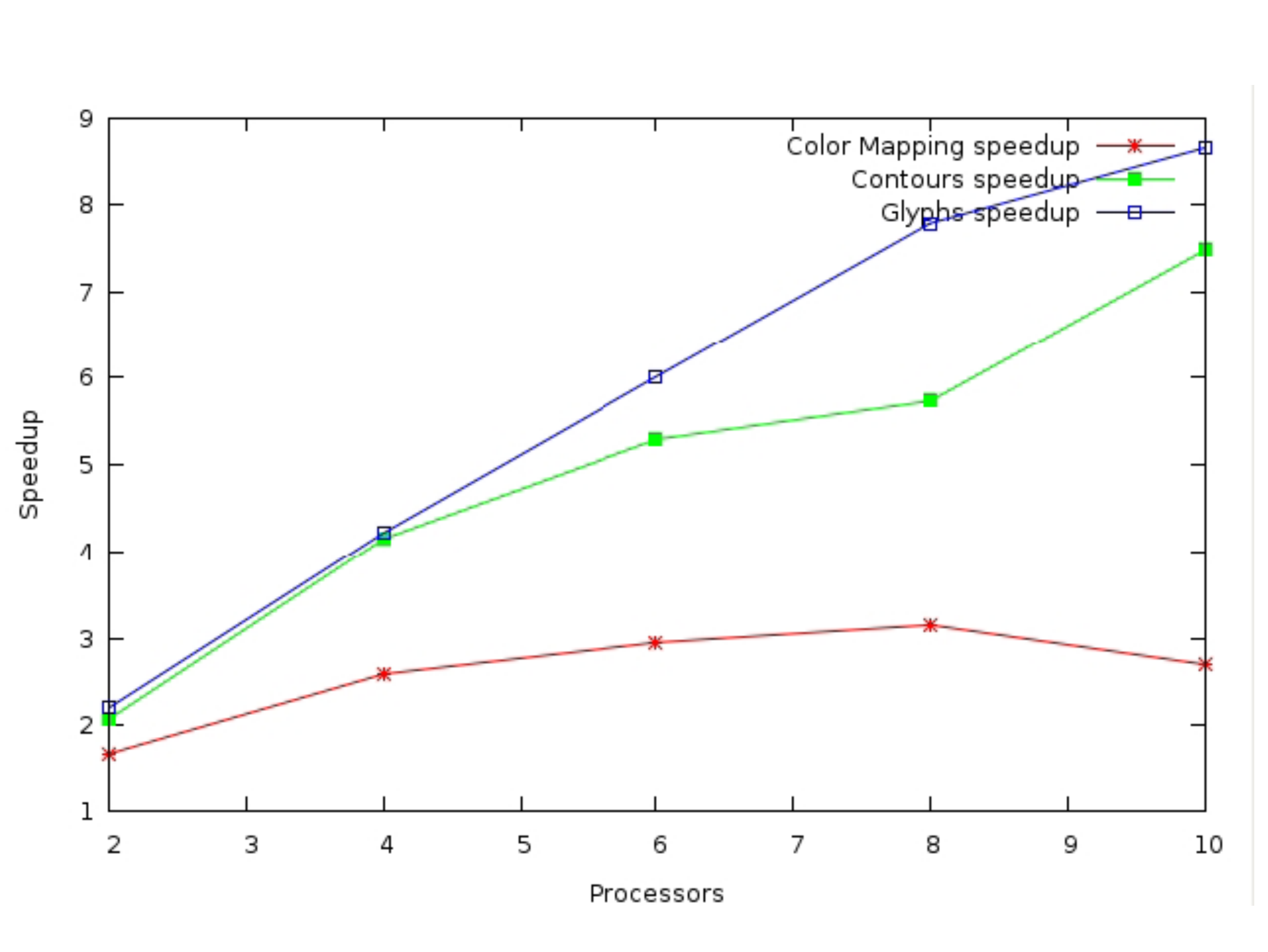}

{\bf (a)}
\end{center}
\end{minipage}
\begin{minipage}{0.49\textwidth}
\begin{center}
\includegraphics[width=\textwidth]{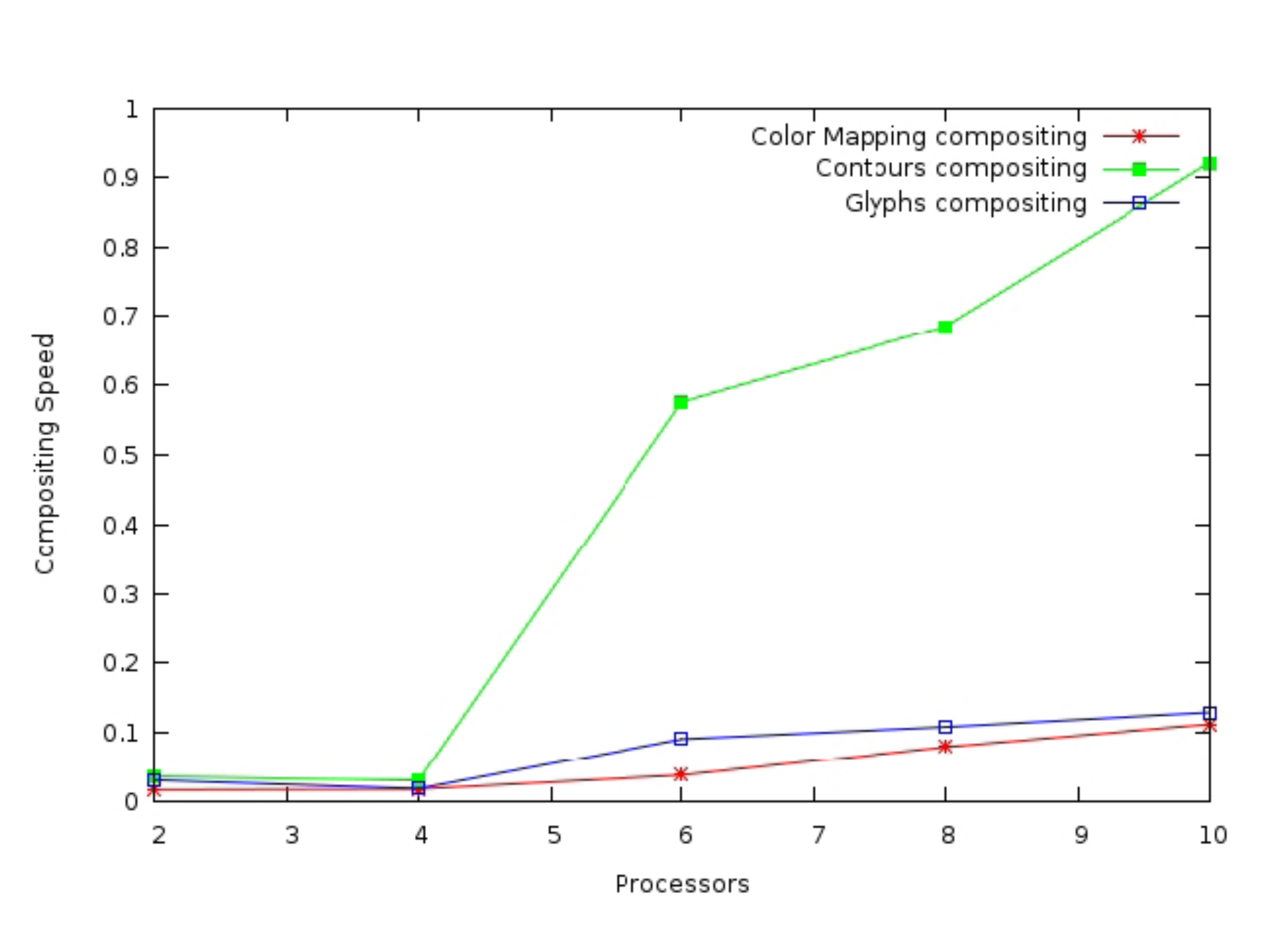}

{\bf (b)}
\end{center}
\end{minipage}
\caption{\label{fig:atanas_03} Speedup evaluation of the parallel visulisation
	 with a domain partitioning defined by the \textsf{Peano} solver and
	 costs for the visualisation composition (taken from
	 \cite{Atanasov:09:DA}). (a) Speedup of the colour mapping, contour
	 generation, and glyphing. (b) Runtime for the composition of
	 pictures.} 
\end{figure}

In case of \textsf{Peano} (see Sect.~\ref{sec:peano}) as simulation code, it is
not neccessary to define a new BSP-tree decomposition of the domain for
visualisation purposes as \textsf{Peano} already provides it for its own domain
decomposition. As this decomposition is already done in a load balanced way
and, in case of a non-\emph{p}-adaptive code such as \textsf{Peano}, simulation
costs as well as visualisation costs per inner domain node are approximately
constant, it can be efficiently used also for parallel visualisation. Test runs
with the steering framework and the CFD solver \textsf{Peano} have been
performed at the Linux Cluster (eight-way AMD Opteron, 2.6~GHz, 32~GByte RAM
per node) at Leibniz Supercomputing Center (LRZ) in Garching.  The
visualisation has been done on a Sun X4600 Server with eight quad-core Opterons
with 256~GByte RAM per processor and four Nvidia Quadro FX5800 graphic cards.
Figure \ref{fig:atanas_03} shows the resulting speedup and the costs for
picture composition. These results are preliminary and still offer a wide range
of optimisation properties both in terms of the number of processors used and
in terms of the speedup.

%
%
\section{Related Applications}
\label{sec:others}
In the following, we will highlight some related applications that have
been developed independent from \textsf{iFluids}. The first one, the
Navier-Stokes solver of the framework \textsf{Peano}, has been the test
application during the development of the steering framework. The second one, a
thermal comfort assessment application, is a steering application not yet
directly related to high-performance computing.  However, to refine the
underlying model -- which will be neccessary in the future -- also fluid
dynamics will have to be included in the model which will than strongly be
related to the main focus of this paper. 

\subsection{Peano}
\label{sec:peano}
\begin{figure}
\begin{minipage}{0.49\textwidth}
\begin{center}
\includegraphics[width=\textwidth]{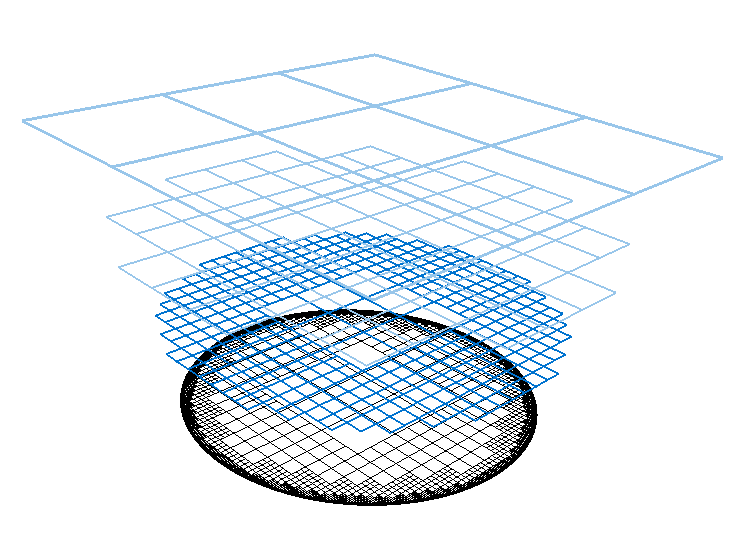}
\end{center}
\end{minipage}
\begin{minipage}{0.49\textwidth}
\begin{center}
\includegraphics[width=0.9\textwidth]{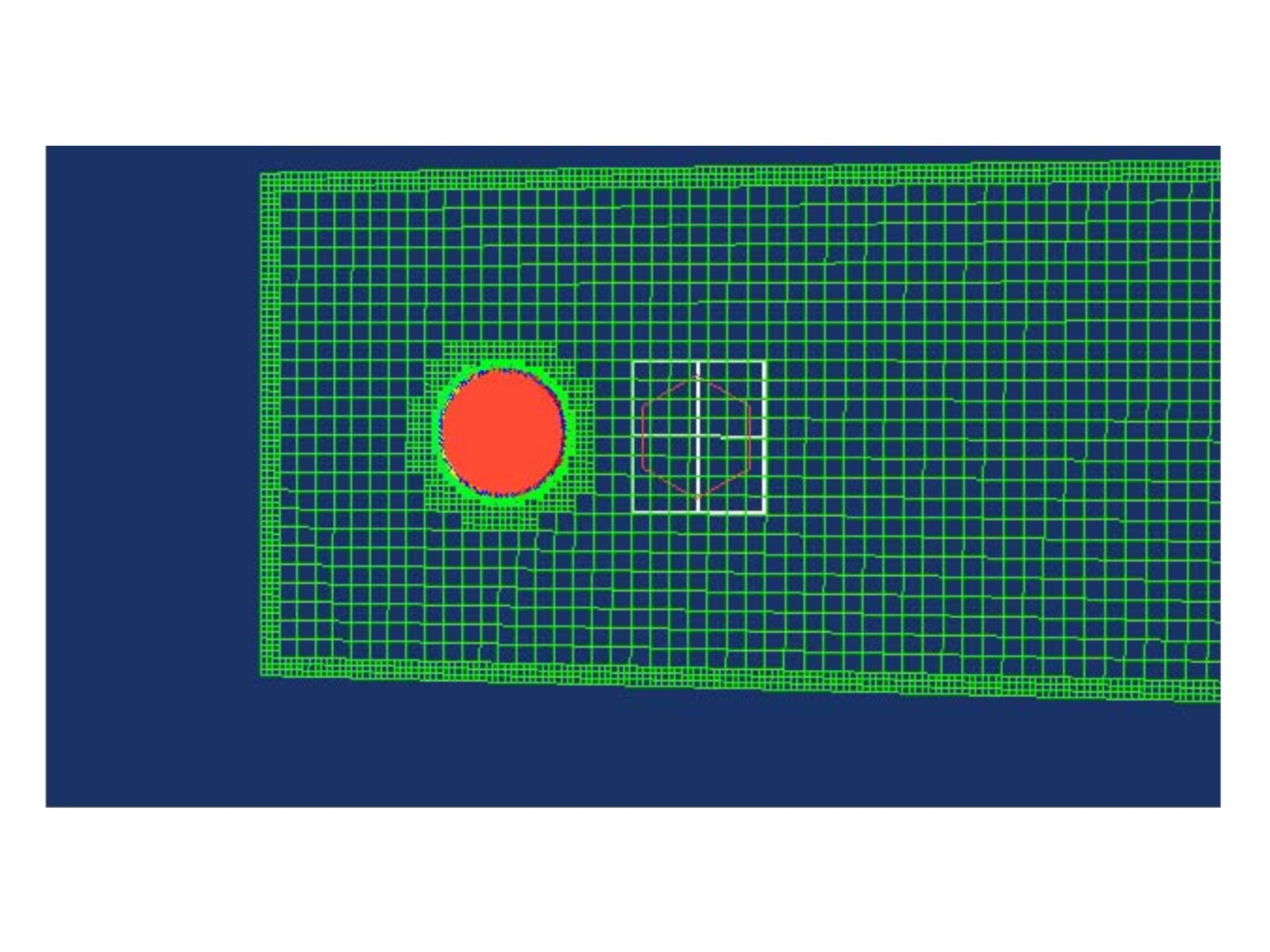}

\vspace{-5mm}
\includegraphics[width=0.9\textwidth]{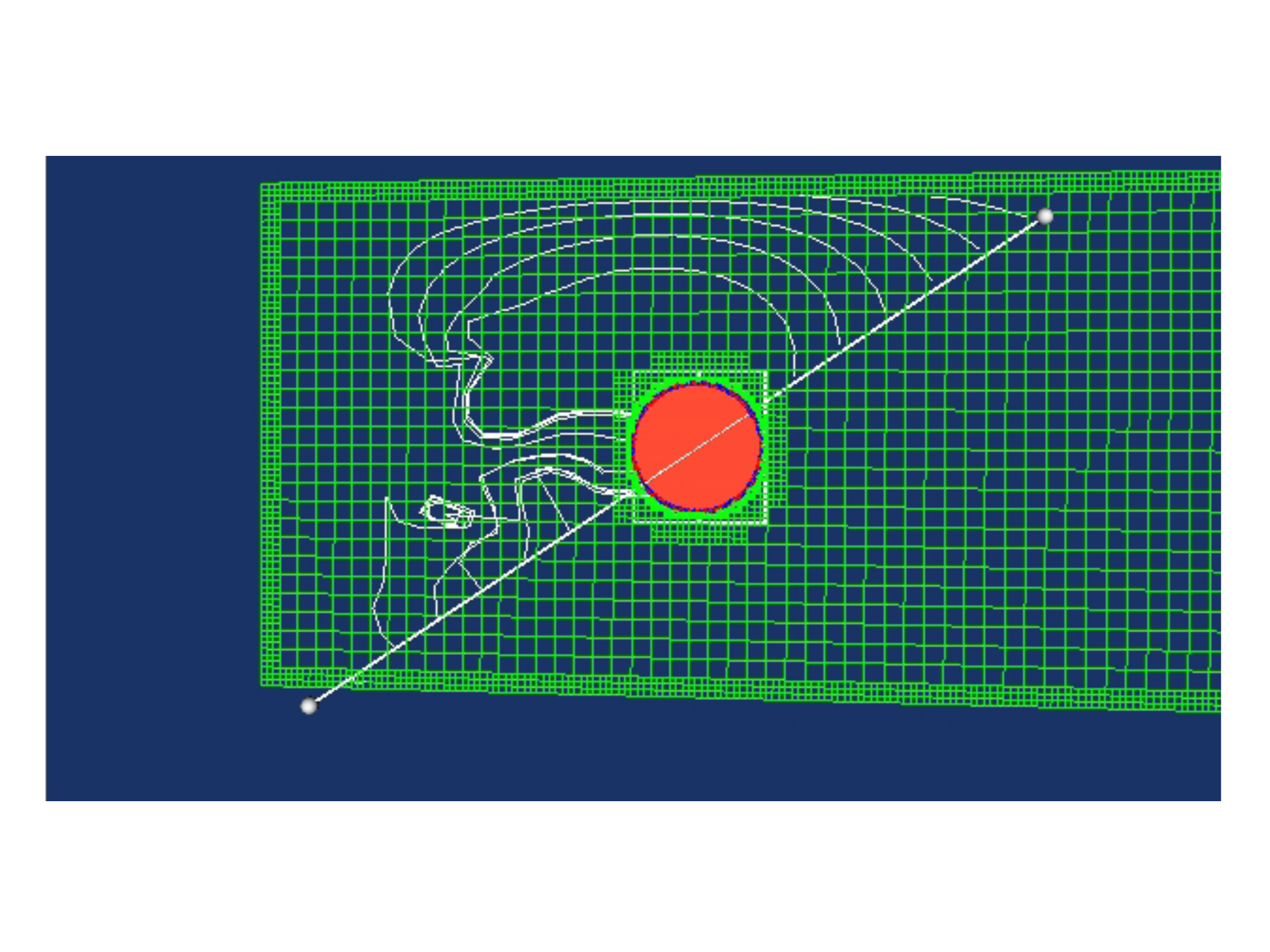}
\end{center}
\end{minipage}

\vspace{-5mm}
\begin{minipage}{0.49\textwidth}
\begin{center}
{\bf (a)}
\end{center}
\end{minipage}
\begin{minipage}{0.49\textwidth}
\begin{center}
{\bf (b)}
\end{center}
\end{minipage}
\caption{\label{fig:peano_grids}\textsf{Peano} grids for two-dimensional
	 examples. (a) Grid hierarchy for a spherical domain; (b) adaptive grid
	 refinement following a moving shpere (taken from
	 \cite{Atanasov:09:DA}).}
\end{figure}

\textsf{Peano} is a solver framework for partial differential equations (PDE)
that works on adaptively refined Cartesian grids corresponding to octree-like
tree structures, so-called space-partitioning grids \cite{Weinzierl:09:Diss}.
Within this framework, a Navier-Stokes solver with dynamical grid refinement is
implemented \cite{Neckel:09:Diss}. This code fits perfectly with the steering
concept described above as it naturally provides the grid hierarchy required
for the hierarchical integration of interactive simulations with large HPC
batch jobs for selected scenarios.  Figure~\ref{fig:peano_grids} (a) shows the
grid hierarchy for a simple two-dimensional example. 

The unique selling points of \textsf{Peano} are low memory requirements in
combination with high cache hit-rates, efficient multiscale solvers, and
efficient and parallel tree-based domain decomposition. \textsf{Peano} has been
run on the HLRB~II at the Leibniz Supercomputing Center in Garching on up to
900 processors with a speedup of 700 \cite{BungartzEtAl:09:PeanoCFD}.  It can
handle moving objects leading to arbitrarily large geometry or even topology
changes as it is based on a fixed (Eulerian) grid. Only the adaptive grid
refinement is adjusted according a deforming, moving, deleted, or added object
(see Fig.~\ref{fig:peano_grids} (b)). Such, also particles advected in a flow
field can be simulated in a very efficient way \cite{Brenk:08:DriftRatchet}. 

Due to its suitability for both the hierarchical integration approach and the
parallel tree-based domain decomposition that can also be used for parallel
visualisation, the test runs for the steering framework described in the
previous section, have been performed with \textsf{Peano} as a simulation code.

\subsection{Thermal Comfort Assessment}

\subsubsection{Motivation}
Indoor climate predictions in office buildings gained increasing importance in
the past. The aim of reducing the energy consumption of buildings, and
maintaining reasonable indoor temperatures for the occupants at the same time,
can be accomplished using simulation tools in the early design stages of the
design phase.

In the broader context of the underlying research project COMFSIM
\cite{Treeck:Bph:07} three modules were defined. In a first study, a
\emph{virtual climate chamber} \cite{vanTreeck:JBPS:2009} was designed, which makes use
of a human thermoregulation model according to Fiala \cite{Fiala:1998}.
Occupants can be situated in a rectangular enclosure with well-defined boundary
conditions, such as room and surface temperatures, relative humidity, air
velocity and metabolic rate. The latter quantities can be changed during an
ongoing simulation using the computational steering concept.

The numerical thermal manikin can be coupled with \textsf{iFluids}
\cite{Treeck:Bph:07}. After a series of iterations of the CFD solver, the
current boundary conditions at the surface of the manikin shall be delivered to
the thermoregulation interface. The existing interface provides the thermal
state of the manikin in terms of the resultant surface temperatures and heat
fluxes, which may act as new boundary conditions of the manikin in the next CFD
step. Using these resulting surface temperatures, a local comfort vote can be
calculated using a 7 point ASHRAE scale \cite{ASHRAE:55:2004}, for example,
indicating the comfort state of the manikin. The developed local assessment
method of our postprocessing tool has already been published by the authors in
\cite{vanTreeck:JBPS:2009}. Coupling CFD with the numerical manikin offers the
possibility to predict the indoor thermal comfort situation in detail, such as
assessing the draught risk, asymmetric radiation, etc. \cite{Treeck:BS:2009}

\subsubsection{Thermoregulation Modeling}
Thermoregulatory reactions of the central nervous system are an answer of
multiple functions of signals from core and peripherals. Local changes in skin
temperature additionally cause local reactions such as modifying the sweating
rate or the local vasodilatation. Significant indicators are the mean skin
temperature and its variation over time and the hypothalamus temperature. The
indicators can be correlated with the autonomic responses in order to form a
detailed thermoregulation model \cite{Fiala:1998,Stolwijk:71}.

Detailed manikin models usually consist of a passive system dealing with
physical and physiological properties, including the blood circulation and an
active thermoregulation system for the afferent signals analysis
\cite{Stolwijk:71}. Local clothing parameters are taken into account and the
response of the metabolism can be simulated over a wide range of ambient
conditions. Besides two-node models (Gagge) \cite{Gagge:1973}, multi-segment
models are known which are founded on the early work of Stolwijk
\cite{Stolwijk:71}. Most models use a decomposition of the human body into
layers and segments for the passive system which are in thermodynamic contact
with each other and with the ambient environment.

As mentioned in section above, the numerical approach for the evaluation of the
human thermoregulation for this application was chosen to be the Fiala model.
Detailed information can be found in \cite{Fiala:1998}.

\subsubsection{Computational Steering Approach}

\begin{figure}[h]
\center
\sidecaption
\includegraphics[width=7.5cm]{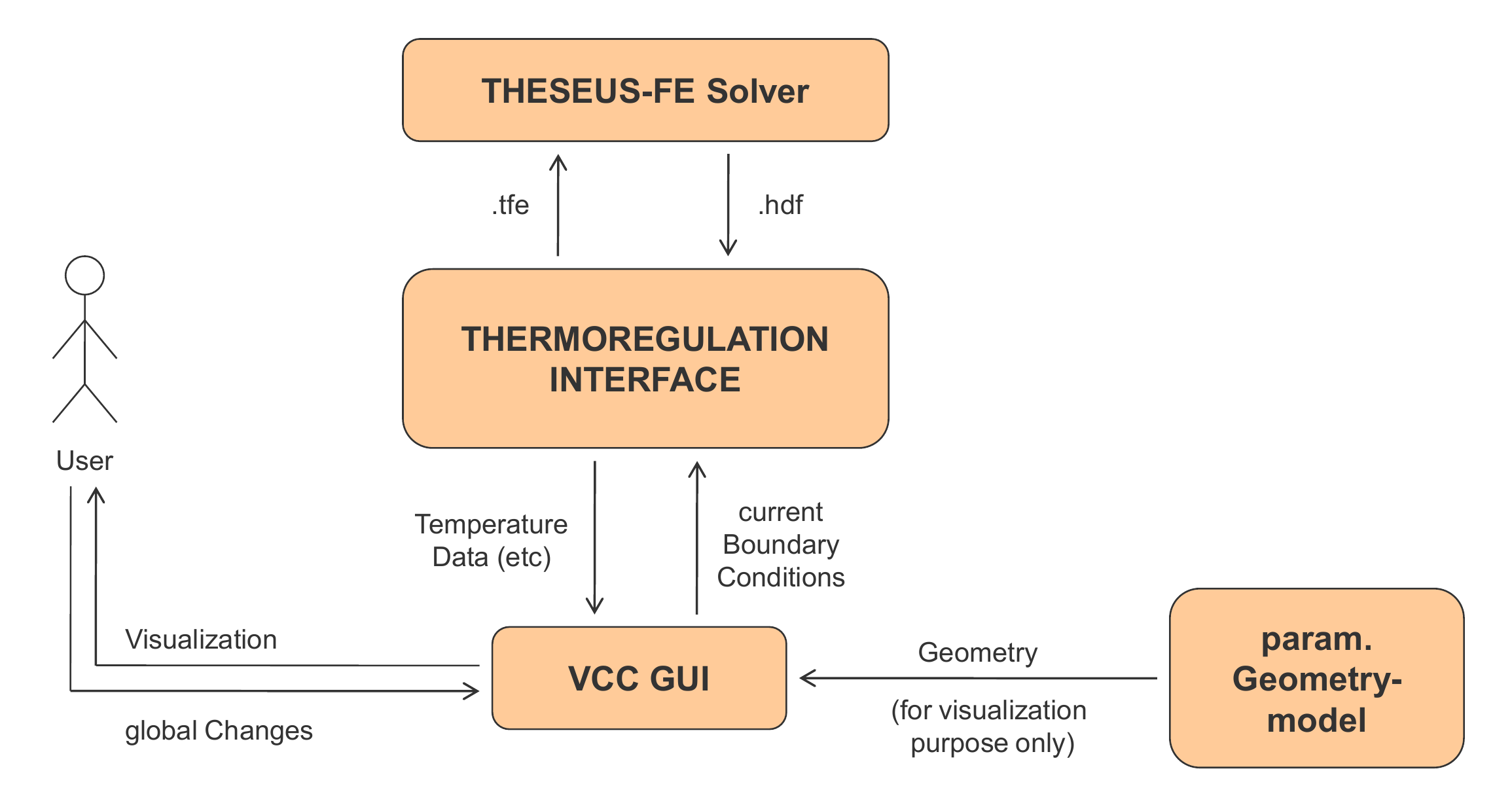}
\caption{Coupling concept: \emph{virtual climate chamber} in computational
         steering mode}
\label{fig:coupl_vcc}
\end{figure}

The above mentioned procedure can be embedded in a computational steering
context. Figure~\ref{fig:coupl_vcc} shows the coupling of the virtual climate
chamber (VCC) with the thermoregulation interface. The user loads the geometry
in to the virtual climate chamber for visualization. There global boundary
conditions can be set, governing the chamber climate. The data is transfered to
the thermoregulation interface which is coupled to a numerical solver. The aim
of the interface is to provide standard interface functions in a way that the
numerical model could be exchanged easily. The numerical model computes a small
timestep and delivers the results to the interface which sends them to the
virtual climate chamber for visualisation purposes. Depending on the just shown
results, the user might want to alter some of the boundary conditions which
will be again transfered to the interface for further treatment and so on.

\begin{figure}[h]
\center
\sidecaption
\includegraphics[width=7.5cm]{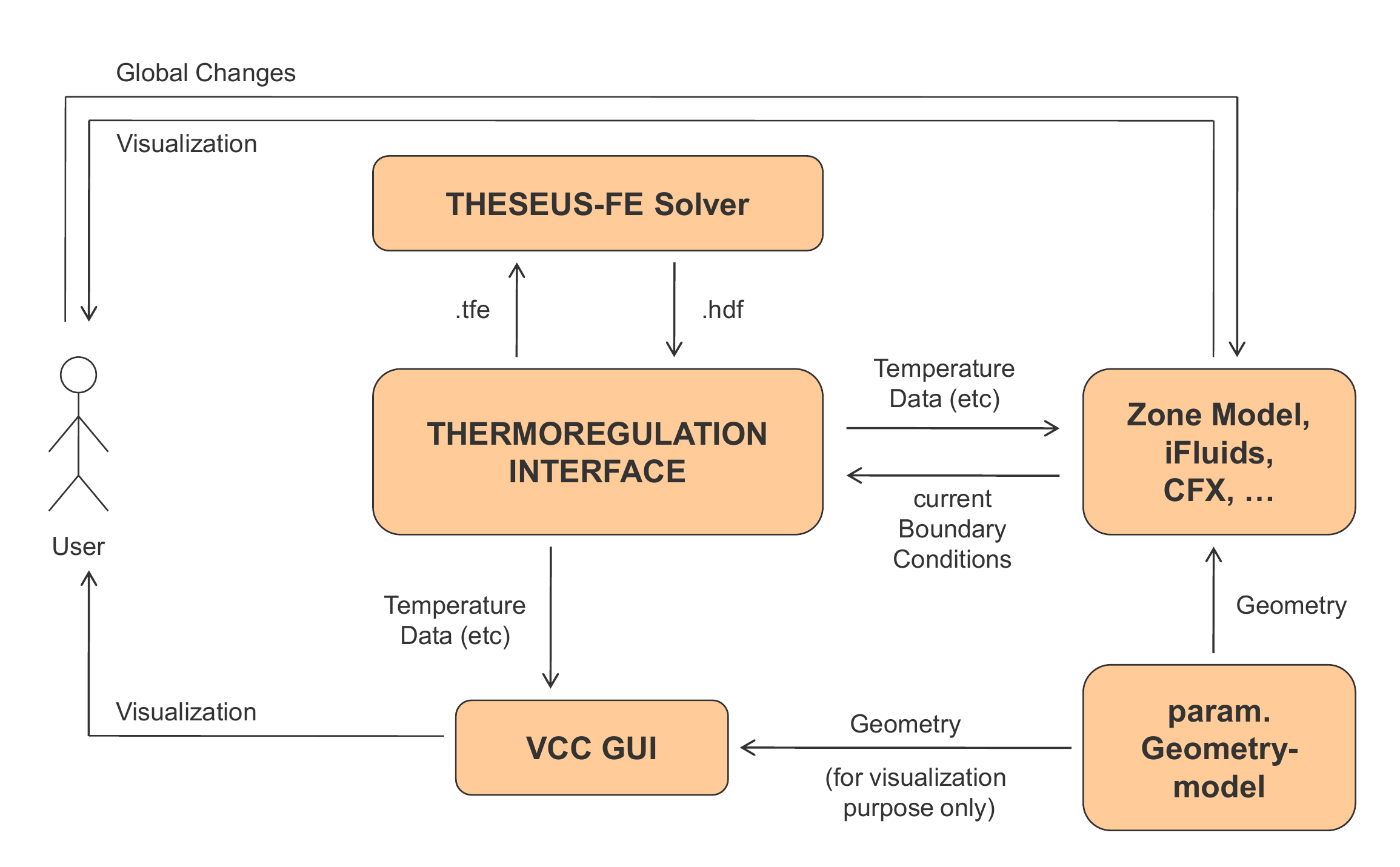}
\caption{Coupling concept: external CFD solvers or zonal models coupled with
         the thermoregulation simulation}
\label{fig:coupl_ifluids}
\end{figure}

This procedure is nice for test cases, but is hardly applicable in real
applications. Therefore a more realistic coupling is depicted in
Fig.~\ref{fig:coupl_ifluids}. The user starts a CFD computation which loads the
geometry and scene information. Manikins are now embedded in the geometry and
classified as \emph{thermal active components}. The CFD code computes a fixed
amount of timestep and delivers the local velocities and temperatures at the
manikin's surface which will be transfered to the thermoregulation interface
who will pass them on to the solver and deliver the results back to the
interface. The resultant surface temperatures are given to the CFD computation
which will act as new boundary conditions in the next CFD step. The
virtual climate chamber is connected to the thermoregulation interface in
\emph{view only} mode in order to observe further detailed information about
the numerical thermoregulation simulation like mean values for the whole body
as mean skin temperature etc.

%
%
\section{Summary and Outlook}
\label{sec:summary}
We proposed tools that combine efficient HPC flow solvers with a steering
environment in order to allow both fast interactive simulations for many
different scenarios and large HPC simulations for selected scenarios in a
hierarchical manner. First tests measuring the performance of the parallel
visualisation tools and the simulation codes on high-performance graphics
hardware and HPC architectures, resp., show promising results.

In the future, the combination of the presented tools shall be applied to
further scenarios and, accordingly, enhanced with more functionality. In
particular, the domain decomposition approach of \textsf{iFluids} will be
improved and particle simulation methods will be implemented in
\textsf{iFluids} and enhanced in \textsf{Peano}. 

%
%
\begin{acknowledgement}
Parts of this work have been carried out with the financial support of KONWIHR
-- the Kompetenznetzwerk f\"ur Technisch-Wissenschaftliches Hoch- und
H\"ochstleistungs\-rechnen in Bayern.
\end{acknowledgement}
%

%
%
\bibliography{bericht}{}
\bibliographystyle{plain}

\end{document}